\documentclass[preprint,aps]{revtex4}
\usepackage{mathrsfs}

\usepackage{graphicx}
\usepackage{multirow}
\usepackage{makecell}
\usepackage{booktabs}

\begin{document}

\title{High Precision Determination of the Planck Constant by Modern Photoemission Spectroscopy}
 \author{Jianwei Huang$^{1,2}$, Dingsong Wu$^{1,2}$, Yongqing Cai$^{1,2}$, Yu Xu$^{1,2}$, Cong Li$^{1,2}$, Qiang Gao$^{1,2}$, Lin Zhao$^{1}$, Guodong Liu$^{1,4}$, Zuyan Xu$^{3}$ and X. J. Zhou$^{1,2,4,5,*}$}

\affiliation{
\\$^{1}$Beijing National Laboratory for Condensed Matter Physics, Institute of Physics, Chinese Academy of Sciences, Beijing 100190, China.
\\$^{2}$University of Chinese Academy of Sciences, Beijing 100049, China.
\\$^{3}$Technical Institute of Physics and Chemistry, Chinese Academy of Sciences, Beijing 100190, China.
\\$^{4}$Songshan Lake Materials Laboratory, Dongguan, Guangdong 523808, China.
\\$^{5}$Beijing Academy of Quantum Information Sciences, Beijing 100193, China
\\$^{*}$Corresponding author: XJZhou@iphy.ac.cn
}


\maketitle

{\bf  The Planck constant, with its mathematical symbol \textit{h}, is a fundamental constant in quantum mechanics that is associated with the quantization of light and matter.   It is also of fundamental importance to metrology, such as the definition of ohm and volt, and the latest definition of  kilogram.  One of the first measurements to determine the Planck constant is based on the photoelectric effect, however, the values thus obtained so far have exhibited a large uncertainty. The accepted value of the Planck constant, 6.62607015$\times$10$^{-34}$ J$\cdot$s,  is obtained from one of the most precise methods, the Kibble balance, which involves quantum Hall effect, Josephson effect and the use of the International Prototype of the Kilogram (IPK) or its copies.  Here we present a precise determination of the Planck constant by modern photoemission spectroscopy technique. Through the direct use of the Einstein's photoelectric equation,  the Planck constant is determined by measuring accurately the energy position of the gold Fermi level using light sources with various photon wavelengths.   The precision of the measured  Planck constant,  6.62610(13)$\times$10$^{-34}$ J$\cdot$s,  is four to five orders of magnitude improved from the previous photoelectric effect measurements. It has rendered photoemission method to become one of the most accurate methods in determining the Planck constant. We propose that this direct method of photoemission spectroscopy has advantages and a potential to further increase its measurement precision of the Planck constant to be comparable to the most accurate methods that are available at present. }

\vspace{10mm}

The Planck constant, \textit{h}, was first proposed by Max Planck more than a hundred years ago when he solved the black body radiation problem by postulating that energy is quantized and can only be emitted or absorbed in integral multiples of a small unit, known as a quantum\cite{1901Planck}. The energy of a particular quantum is described  by the equation $E$ = $h\nu$ where $\nu$ represents the frequency of the radiation and \textit{h} is  the Planck constant.  Subsequently in 1905,  Albert Einstein extended Planck's black body model to explain the photoelectric effect by describing light as composed of discrete quanta called photons with an energy of $h\nu$ where  $\nu$ represents the frequency of incident light\cite{1905EinsteinPhoton}.  The photoelectric effect is interpreted in terms of the Einstein's photoelectric equation: $h\nu$ = $\Phi$ + $E_{max}$ where  $\Phi$  is the work function of the metal and $E_{max}$ is the maximum kinetic energy of the ejected electron(s). Since then, the Planck constant has played a major role in giving birth to quantum physics\cite{1924deBroglie,1981PAMDirac,1927WHeisenberg} and has become one of the most important universal constants in physics.

The precise determination of the Planck constant is also of fundamental importance to metrology. The SI standard of resistance, ohm, is related to the von Klitzing constant $R_K$ = $h/e^2$ from quantum Hall effect\cite{2005vonKlitzing}. Here \textit{h} is  the Planck constant and \textit{e} is the elementary charge. The SI standard of voltage, volt,  is related to the Josephson constant $K_J$ = $2e/h$ from Josephson effect\cite{2000VoltReview}.    The latest definition of mass, kilogram,  is based on taking the fixed numerical value of the Planck constant to be  6.62607015$\times$10$^{−34}$ J$\cdot$s\cite{KiloDefinition}.  The value of the Planck constant was first determined from fitting the black body radiation curve\cite{1901Planck}. Such a method is straightforward but was found difficult to produce a reproducible high precision result\cite{hfromBBR1,hfromBBR2,hfromBBR3,hfromBBR4}. The photoelectric effect provides another way to measure the Planck constant, however, the value of the Planck constant thus obtained so far varies over a large range\cite{1912ComptonKT,1913HughesAL,1914RichardsonOW,1916MillikanRA,HallHH1971TuttleRP,BoysDW1978MykolajenkoW,BarnettJD1988StokesHT,LoparcoF2017SpinelliP}. At present, there are a couple of methods that have been invented to precisely determine the Planck constant\cite{KibbleBP1979HuntGJ,BowerVE1980DavisRS,ClothierWK1989BenjaminDJ,FunckT1991SienknechtV,FujiiK2005HankeM}.  One of the most precise methods is based on the Kibble balance which can determine the Planck constant to a precision that is less than 2$\times$10$^{-8}$ relative uncertainty\cite{SanchezCA2014InglisD,SchlammingerS2014PrattJR,RobinsonIA2016SchlammingerS,WoodBM2017LiardJO}.  However, in order  to conduct the measurement to achieve such a high precision, complex apparatus  has to be constructed that involves the Josephson effect, the quantum Hall effect and the International Prototype of the Kilogram (IPK) or its copies\cite{SanchezCA2014InglisD,SchlammingerS2014PrattJR,RobinsonIA2016SchlammingerS,WoodBM2017LiardJO}.


In principle, the photoelectric effect provides a simple and direct way to measure the Planck constant.  From the Einstein's photoelectric equation: $h\nu$ = $\Phi$ + $E_{max}$,  the maximum photoelectron energy $E_{max}$ is proportional to the light wave frequency $\nu$, and the ratio is the Planck constant \textit{h}.  Millikan proved the linear relation between $E_{max}$ and $\nu$ with his state-of-the-art apparatus at that time\cite{1916MillikanRA} and many photoelectric effect measurements\cite{HallHH1971TuttleRP,BoysDW1978MykolajenkoW,BarnettJD1988StokesHT,LoparcoF2017SpinelliP} adopted his method. An extra potential was applied at the electron collector to stop the photoelectron from arriving at the collector and the maximum photoelectron energy $E_{max}$ is determined in this case by measuring the electric current flowing through the collector\cite{1916MillikanRA}. The main difficulty and source of uncertainty originate from how to define the stop potential and determine it accurately from the photocurrent-potential curve\cite{1916MillikanRA,LoparcoF2017SpinelliP}. Since electrons in solids obey the Fermi-Dirac distribution around the Fermi level, there is actually no clear definition of the maximum photoelectron energy at a finite temperature. In fact,  $E_{max}$ in the photoelectric effect equation represents the maximum photoelectron energy excited from a metal at zero temperature.  Another source of uncertainty comes from the absorption of residual gas on the clean metal surface; it will affect the stability of the photocurrent and cause variations in the work function and the stop potential. The stability of the light source is also important for the precise measurement of the photocurrent. Overall, these factors have made it hard to determine $E_{max}$ from the photocurrent-potential curve with high precision,  giving rise to  a rather large uncertainty in  the value of the Planck constant obtained from the photoelectric effect\cite{1912ComptonKT,1913HughesAL,1914RichardsonOW,1916MillikanRA,HallHH1971TuttleRP,BoysDW1978MykolajenkoW,BarnettJD1988StokesHT,LoparcoF2017SpinelliP}.

The photoemission spectroscopy technique has experienced a dramatic advancement in the last few decades\cite{HuefnerBook,ADamascelli2003ZXShen,GDLiu2008XJZhou,XJZhouReview}. In modern photoemission spectroscopy, the energy of the photoelectrons can be precisely measured, and the emission angle of photoelectrons can also be measured that provides information of electron momentum in the measured material.  The light source is improved and the measured sample can be kept in ultra-high vacuum to stay clean and stable.  These can overcome the issues encountered by the previous photoelectric effect measurements\cite{1912ComptonKT,1913HughesAL,1914RichardsonOW,1916MillikanRA,HallHH1971TuttleRP,BoysDW1978MykolajenkoW,BarnettJD1988StokesHT,LoparcoF2017SpinelliP} and prompt us to ask whether the Planck constant can be determined precisely by using modern photoemission technique. In this paper, we report a high precision determination of the Planck constant by employing modern photoemission spectroscopy technique. Through the use of the Einstein's photoelectric equation,  the Planck constant is directly determined by measuring accurately the energy position of the gold Fermi level using light sources with various photon wavelengths.   The precision of the measured  Planck constant,  6.62610(13)$\times$10$^{-34}$ J$\cdot$s,  is four to five orders of magnitude improved from the previous photoelectric effect measurements, and renders photoemission method to become one of the most accurate methods in determining the Planck constant. With further optimization of the photoemission apparatus,  we believe this direct method of photoemission spectroscopy is possible to reach its measurement precision of the Planck constant to the level that is comparable to the most accurate methods that are available at present.

A typical setup of modern photoemission system, angle-resolved photoemission spectroscopy (ARPES), is schematically shown in Fig. 1a. It is still based on the photoelectric effect. When light is incident on the sample in an ultra-high vacuum chamber, electrons in the sample absorb the photons and photoelectrons are emitted outside of the sample. By measuring the energy and number of the photoelectrons along different emission angles, one may get the electronic structure of the sample in terms of electron energy and momentum. One key element to measure the energy, number and emission angle of the photoelectrons is the electron energy analyzer.  In Fig. 1a, a typical and most commonly used hemispherical electron energy analyzer is shown which consists of a lens system, a slit, the inner and outer spheres, and an electron detector. Electrons entering through the same position of the slit with different energies will deflect with different radius in the hemispherical analyzer because of the application of a potential difference on the outer and inner spheres. The photoelectrons with different energies are dispersed along the vertical direction of the detector at the exit of the hemispheres.  The intensity of photoelectrons dispersed at different energies is measured at different vertical locations of the detector. The hemispherical analyzer can work in two different modes, one is transmission mode and the other is angular mode.  In the transmission mode (Fig. 1c), the photoelectrons with different emission angles at the same spot on the sample will be focused on the same point on the slit, giving rise to an angle-integrated photoemission spectrum. In the angular mode  (Fig. 1d), the photoelectrons with different emission angles at the same spot on the sample will be dispersed by the lens system to different horizontal positions on the slit, and further spread along the horizontal direction on the detector, giving rise to an angle-resolved photoemission image.

High resolution photoemission measurements were performed using a lab-based ARPES system equipped with a helium discharge lamp and a hemispherical electron energy analyzer\cite{GDLiu2008XJZhou}.  A monochromator is used to achieve three kinds of monochromatic light with the wavelength $\lambda$ of 58.43339 nm (He I$\alpha$), 53.70293 nm (He I$\beta$) and 30.37858 nm (He II$\alpha$). The accurate wavelength of the light  is obtained from the NIST Atomic Spectra Database where  the wavelength of He I$\alpha$ and $\beta$ is obtained by experimental measurements and the wavelength of He II$\alpha$ is obtained by theoretical calculation\cite{NISTatomicspectrawebsite}. The energy resolution of the analyzer was set at 2.5 meV and the angular resolution was $\sim$0.3$^o$. We use a polycrystalline gold foil as the sample of the photoelectron emission  because it can provide a well-defined Fermi edge that is an intrinsic physical quantity which can be described by the Fermi-Dirac distribution.  The advantage of using the gold Fermi edge in defining a precise energy position is obvious when compared with the maximum energy of the photoelectrons $E_{max}$  used in the previous photoelectric measurements\cite{1912ComptonKT,1913HughesAL,1914RichardsonOW,1916MillikanRA,HallHH1971TuttleRP,BoysDW1978MykolajenkoW,BarnettJD1988StokesHT,LoparcoF2017SpinelliP}.  The polycrystalline gold was sputtered by an Argon ion gun to get a clean surface before it was transferred to the ultra-high vacuum chamber with a base pressure better than 5$\times$10$^{-11}$ mbar. Such an ultra-high vacuum keeps the sample surface clean and stable during the measurement.   In order to get a sharp Fermi cut-off to determine its energy position accurately,  the polycrystalline gold was kept at a temperature of 1.4 K that was precisely controlled to be within $\pm$0.1 K stability.

In our present measurement,  we used a polycrystalline gold metal as the target material and measured the kinetic energy of the photoemitted electrons after being excited by light. The polycrystalline gold is electrically connected with the electron energy analyzer and both are grounded (Fig. 1a). Therefore, the Fermi level of the polycrystalline gold and the analyzer are lined up at the same energy level due to the good electrical contact (Fig. 1b). The vacuum level can be different because the work function (energy difference between the Fermi level and the vacuum level) of the polycrystalline gold ($\Phi_{Au}$) and the analyzer ($\Phi_{Ana}$) can be different.  When the light with a photon energy of $h\nu$ is incident on the polycrystalline gold, the kinetic energy of the photoelectrons at the Fermi level is $E^{\prime}_{KF}$ = $h\nu - \Phi_{Au}$ according to the Einstein's photoelectric equation.  These photoemitted electrons, upon entering the analyzer lens system from the gold sample surface,  will experience the potential difference between the gold  and the analyzer that is ($\Phi_{Au} - \Phi_{Ana}$) to get accelerated or decelerated depending on the relative magnitude of the work function between Au and the analyzer. The  measured kinetic energy of the photoelectrons at the Au Fermi level, after exiting the electron energy analyzer, is then $E_{KF}$ = $h\nu - \Phi_{Au} + (\Phi_{Au} - \Phi_{Ana}$) = $h\nu - \Phi_{Ana}$ = $hc/\lambda - \Phi_{Ana}$ where $c$ is the speed of light.  It is independent on the Au work function but related to the work function of the analyzer that is a constant within the measurement frame of time. From this modified version of the Einstein's photoelectric equation, when the energy position of the Au Fermi level ($E_{KF}$) is determined using light with different wavelengths $\lambda$, the Planck constant can be determined directly from the slope of the linear relation.

Figure 2 shows the measured Au Fermi edge using light with different wavelengths under different measurement modes of the electron energy analyzer.  Figure 2(a-c) shows the Au Fermi edge measured at a temperature of 1.4 K by transmission mode using light with three wavelengths of 58.43339 nm (He I$\alpha$), 53.70293 nm (He I$\beta$) and 30.37858 nm (He II$\alpha$), respectively. In this case, photoelectrons within an emission angle of $\sim \pm$15 degrees with respect to the sample normal are collected to give one photoemission spectrum (energy distribution curve, EDC) for each photon wavelength. The horizontal axis is the measured photoelectron kinetic energy while the vertical axis is the intensity of photoelectrons detected.  The EDC of the measured Au Fermi edge can be described by the Fermi-Dirac distribution function convolved with the instrumental energy resolution with the formula:  $f(\varepsilon)\otimes R(\bigtriangleup E) = 1/(e^{(\varepsilon - E_{KF})/(k_BT)} + 1) \otimes e^{-\varepsilon^2/(2\bigtriangleup E^2)}$ where $\varepsilon$ is the photoelectron energy and $\bigtriangleup E$ is the instrumental energy resolution.  Precise values of $E_{KF}$ can be obtained by fitting the EDCs of the measured Au Fermi edge with the formula and the obtained values are shown in Fig. 2a, b and c with the fitting uncertainty in the parentheses. We calculate their relative uncertainty as 6.5$\times$10$^{-6}$, 13.4$\times$10$^{-6}$ and 7.1$\times$10$^{-6}$ for 58.43339 nm (He I$\alpha$), 53.70293 nm (He I$\beta$) and 30.37858 nm (He II$\alpha$) light, respectively.

Figure 2(d-f)  shows the measured photoemission images of the Au Fermi edge at 1.4 K by the angular mode of the electron energy analyzer using light with three different photon wavelengths. The horizontal axis of the images is photoelectron kinetic energy and the vertical axis represents the emission angle of these photoelectrons with respect to the Au surface normal (during the measurement, the Au was put with its surface normal along the lens axis of the electron energy analyzer). The false colors in the images represent the photoelectron intensity. In the angular mode, photoelectrons with different emission angles can be separated and measured in parallel, different from the transmission mode where the photoelectrons with different emission angles are measured in sum.  Photoemission spectra (EDCs) can be obtained for each emission angle, or for a range of the emission angle from these measured images. To enhance the data statistics, here we chose to integrate the emission angle between -6$^\circ$ and 6$^\circ$ to get a single integrated EDC (Fig. 2g, h and i).  We note that there is a balance between the data statistics and the accuracy of $E_{KF}$.  A large integration angle range is good for improving the data statistics, but in the mean time may  reduce the accuracy of  $E_{KF}$ because the instrumental aberration increases when the measured channels move away from the central lens and detection regions (corresponding to zero emission angle).  In this sense, the angular mode measurements are advantageous over the transmission mode because the latter integrates photoelectrons over a much larger emission angle ($\pm$15$^\circ$).  Similar to the transmission data in Fig. 2(a-c), the integrated EDCs in Fig. 2(g-i) are fitted and the obtained  $E_{KF}$  is shown in each panel with the fitting uncertainty in the parentheses following each value.  Their relative uncertainty is calculated as 7.7$\times$10$^{-6}$, 16.0$\times$10$^{-6}$ and 7.7$\times$10$^{-6}$ for 58.43339 nm, 53.70293 nm and 30.37858 nm light, respectively.

In Fig. 3, the obtained $E_{KF}$ is plotted as a function of the wave number of the incident light. Here the wave number represents the inverse of the wavelength for each light source. The measured results from the transmission mode (Fig. 2(a-c)) are shown in Fig. 3a, and from the angular mode (Fig. 2(g-i)) are shown in Fig. 3b. In order to check on the effect of the integration angle range on the results, we also used different angle integration windows of [-5,5], [-4,4] and [-3,3] degrees from the measured images (Fig. 2(d-f)), in addition to the [-6,6] degree used in Fig. 2(g-i). The obtained $E_{KF}$s as a function of the wave number of the incident light sources are shown in Fig. 3(c-e).  The data in each panel is fitted with a linear function $\emph{y} = a\emph{x} + b$. According to the modified Einstein's photoelectric equation: $E_{KF}$  = $hc/\lambda - \Phi_{Ana}$,   the fitted slope $a$ corresponds to $hc/e$ where $c$ is the speed of light (299792458 m/s) and $e$ is the elementary charge (1.6021766208$\times$10$^{-19}$ C).  The absolute value of the fitted intercept $b$  corresponds to the work function of the electron energy analyzer ($\Phi_{Ana}$). From the transmission mode data (Fig. 3a), the fitted slope  $a$ is 1.23985719$\times$10$^{-4}$ eV$\cdot$cm and the obtained Planck constant equals to  6.62615138$\times$10$^{-34}$ J$\cdot$s. The fitted work function of the analyzer is 4.3621 eV. These values are marked in the panel of Fig. 3a. The Planck constant thus obtained from the angular mode measurements with different angle integration windows are also marked in Fig. 3(b-e).

Figure 3f summarizes the measured Planck constant values from the transmission mode and angular mode measurements with different angle integration windows from Fig. 3(a-e). In order to further check the effect of the selected energy range on the results in fitting the measured Fermi edges using the Fermi-Dirac distribution function, different energy windows, [-30,30], [-40,40] and [-50,50] meV with respect to the Fermi level position $E_{KF}$, were also tested in fitting the integrated EDCs from the measured images of Fig. 2(d-f). By considering both the transmission mode and angular mode measurements, the effect  of different angle integration windows, and the effect of different fitting energy windows,  Fig. 3f  covers all the obtained Planck constant values.  The averaged value  is 6.62609677$\times$10$^{-34}$ J$\cdot$s with a relative uncertainty of 2$\times$10$^{-5}$. We can write our measured \textit{h} as 6.62610(13)$\times$10$^{-34}$ J$\cdot$s. Taking the accepted Planck constant value \textit{h} = 6.62607015$\times$10$^{-34}$ J$\cdot$s as a reference,  the maximum relative deviation is below 1.6$\times$10$^{-5}$.

In Fig. 4, we compare our measured values of the Planck constant with those from the previous measurements.  When compared with the previous values of the Planck constant obtained based on the photoelectric effect (Fig. 4a)\cite{1912HughesAL,RichardsonOW1912ComptonKT,1916MillikanRA,LukirskyP1928PrilezaevS,HughesAL1932DuBridgeLA,1946OlearyAJ,HallHH1971TuttleRP,BoysDW1978MykolajenkoW,BarnettJD1988StokesHT,LoparcoF2017SpinelliP}, our result is 4$\sim$5 orders of magnitude increased in precision. This is mainly due to the significant advancement in precise determination of the energy scale of the well-defined Au Fermi edge.  When compared with the most accurate methods that have been invented so far (Fig. 4b)\cite{KibbleBP1979HuntGJ,BowerVE1980DavisRS,ClothierWK1989BenjaminDJ,FunckT1991SienknechtV,FujiiK2005HankeM,SanchezCA2014InglisD,SchlammingerS2014PrattJR,RobinsonIA2016SchlammingerS,WoodBM2017LiardJO}, our result has made the measured Planck constant comparable to the most precise results, putting the photoelectric effect method back as one of the most accurate methods in determining the Planck constant.

The present results are obtained from a regular lab-based ARPES system without any particular modification or optimization made for the purpose of the Planck constant measurement.  The electron energy analyzer we used is a commercial product without high precision control of the voltages on the inner sphere, outer sphere, and in particular the bias to tune the energy of the photoelectrons before they pass through the slit.  We also note that, although the achieved precision of our present measurement, 2.0$\times$10$^{-5}$,  is lower than those two of the most accurate methods: Kibble balance method (3.4$\times$10$^{-8}$)\cite{SanchezCA2014InglisD,SchlammingerS2014PrattJR,RobinsonIA2016SchlammingerS,WoodBM2017LiardJO} and the X-ray crystal density method (2.9$\times$10$^{-7}$)\cite{FujiiK2005HankeM}, there remains a lot of room to improve the present photoemission system to further increase its precision. As shown in the modified Einstein's photoelectric equation,  $E_{KF}$ = $h\nu - \Phi_{Ana}$, the modern photoemission method is straightforward with inherent advantages. The precision of the Planck constant measurement in this method relies on two parameters: the light wavelength and the energy position of the Au Fermi edge.  In our present measurements, the relative uncertainty of the wavelength is 0.9$\times$10$^{-6}$, 0.9$\times$10$^{-6}$ and 0.6$\times$10$^{-9}$ for the three light sources 58.43339 nm, 53.70293 nm and 30.37858 nm, respectively.  The light wavelength can be measured to a higher precision of 10$^{-8}$\cite{Wavelength}.  The precise energy position determination of  the Au Fermi edge depends on two factors.  The first is the bias voltage that is used to vary the photoelectron kinetic energy just before they enter the hemisphere through the slit. With the latest technology, it is possible to control the precision and stability of the bias voltage to the level of a micro-volt ($\mu$V).  The second factor is the determination of the measured Au Fermi edge; its precision depends on the measured EDC lineshape, its transition width and the overall data statistics.  If the polycrystalline gold can be cooled to a very low temperature $\sim$1 K (corresponds to a Fermi edge width of 330 $\mu$eV), and the data are taken with high system stability and high data statistics, it is possible to measure the Fermi edge to a precision of $\mu$eV.  Overall, when  the combined accuracy of the energy position for the Au Fermi edge from different photon sources is controlled to be at the $\mu$eV level over a span of the photon energy of $\sim$20 eV between 21.2 eV (He I$\alpha$) and 40.8 eV (He II$\alpha$), the precision of $E_{KF}$ can also approach 10$^{-8}$ level. These would make the modern photoemission method possible to measure the Planck constant to the precision of $\sim$10$^{-8}$ level that is comparable to the most accurate methods that are available so far\cite{SanchezCA2014InglisD,SchlammingerS2014PrattJR,RobinsonIA2016SchlammingerS,WoodBM2017LiardJO,FujiiK2005HankeM}.


The determination of the Planck constant with high precision is of paramount importance to both the metrology and quantum physics. First,  it remains to be checked on the inconsistency of the measured values between the Kibble balance method and the X-ray crystal density method, the two most accurate measurement methods of the Planck constant. The Kibble balance method gives a value of \textit{h} = 6.62606889(29)$\times$10$^{-34}$ J$\cdot$s with a precision of 3.4$\times$10$^{-8}$\cite{SanchezCA2014InglisD,SchlammingerS2014PrattJR,RobinsonIA2016SchlammingerS,WoodBM2017LiardJO}.  The X-ray crystal density method gives a value of \textit{h} = 6.6260745(19)$\times$10$^{-34}$ J$\cdot$s with a precision of 2.9$\times$10$^{-7}$\cite{FujiiK2005HankeM}.  However, the two values of the Planck constant appear not to agree with each other within their claimed precision. Independent measurement from the third method is necessary to resolve this discrepancy. Second, there are several related constants, like the Josephson constant ($K_J$ = $2e/h$) and von Klitzing constant ($R_K$ = $h/e^2$)  that are related to the Planck constant. The precision improvement of the Planck constant will help elevate the precision of other constants. Third, all the precise methods of the Planck constant, except for the X-ray crystal density method, rely on the theoretical basis of the Josephson effect and the quantum Hall effect. In particular, the Kibble balance method involves the use of the International Prototype of the Kilogram (IPK) or its copies, in addition to the involvement of the Josephson effect and the quantum Hall effect to conduct the measurement\cite{SanchezCA2014InglisD,SchlammingerS2014PrattJR,RobinsonIA2016SchlammingerS,WoodBM2017LiardJO}. The values of the Planck constant obtained in this way cannot be used as tests of the theories without falling into a circular argument. An independent measurement with a comparable precision would be important to examine on the accuracy of the measured value, validity of the related theories and possible time evolution of some fundamental constants\cite{2005vonKlitzing}.

In summary, we have measured the Planck constant by conducting the photoelectric effect experiment with the modern photoemission technique. From our lab-based system, we have  obtained the value of the Planck constant  6.62610(13)$\times$10$^{-34}$ J$\cdot$s with a relative uncertainty of 2$\times$10$^{-5}$. The precision is 4 or 5 orders of magnitude improved compared with all the previous measurements based on the photoelectric effect, and puts the technique into the category as one of the most accurate methods in measuring the Planck constant.  The photoelectric effect method  is direct and intuitive with inherent advantages.  We propose that there is still a lot of room to further improve the photoemission technique to achieve a precision that is comparable to the Kibble balance or other precise methods. To this end, a dedicated photoemission system with all the elements optimized, including the light source, the target sample, and the electron energy analyzer,  is desired. We hope our present work will stimulate further efforts along this direction. It provides an opportunity to provide an independent high precision measurement of the Planck constant that will not only check on the other methods, but provide possibility in elevating precision of other fundamental constants and cross-examining some theories including the Josephson effect and quantum Hall effect.



\vspace{3mm}


\vspace{3mm}

\noindent {\bf Acknowledgement}  This work is supported by the National Key Research and Development Program of China (Grant No. 2016YFA0300300 and 2017YFA0302900), the National Natural Science Foundation of China  (Grant No. 11888101),  the Strategic Priority Research Program (B) of the Chinese Academy of Sciences (XDB25000000),   and the Research Program of Beijing Academy of Quantum Information Sciences (Grant No. Y18G06).

\vspace{3mm}

\noindent {\bf Author contributions}\\
 X.J.Z. and J.W.H. proposed and designed the research. J.W.H., D.S.W., Y.Q.C., Y.X., C.L., Q.G., L.Z., G.D.L., Z.Y.X. and X.J.Z. contributed to the development and maintenance of the ARPES system. J.W.H. carried out the ARPES experiment with D.S.W. and Y.Q.C.. J.W.H. and X.J.Z. analyzed the data. J.W.H. and X.J.Z. wrote the paper. All authors participated in discussion and comment on the paper.\\

\vspace{3mm}

\noindent{\bf Additional information}\\
Correspondence and requests for materials should be addressed to X.J.Z.

\vspace{3mm}

\noindent{\bf Competing interests:} 
The authors declare no competing interests.

\newpage

\begin{figure*}[tbp]
\begin{center}
\includegraphics[width=0.8\columnwidth,angle=0]{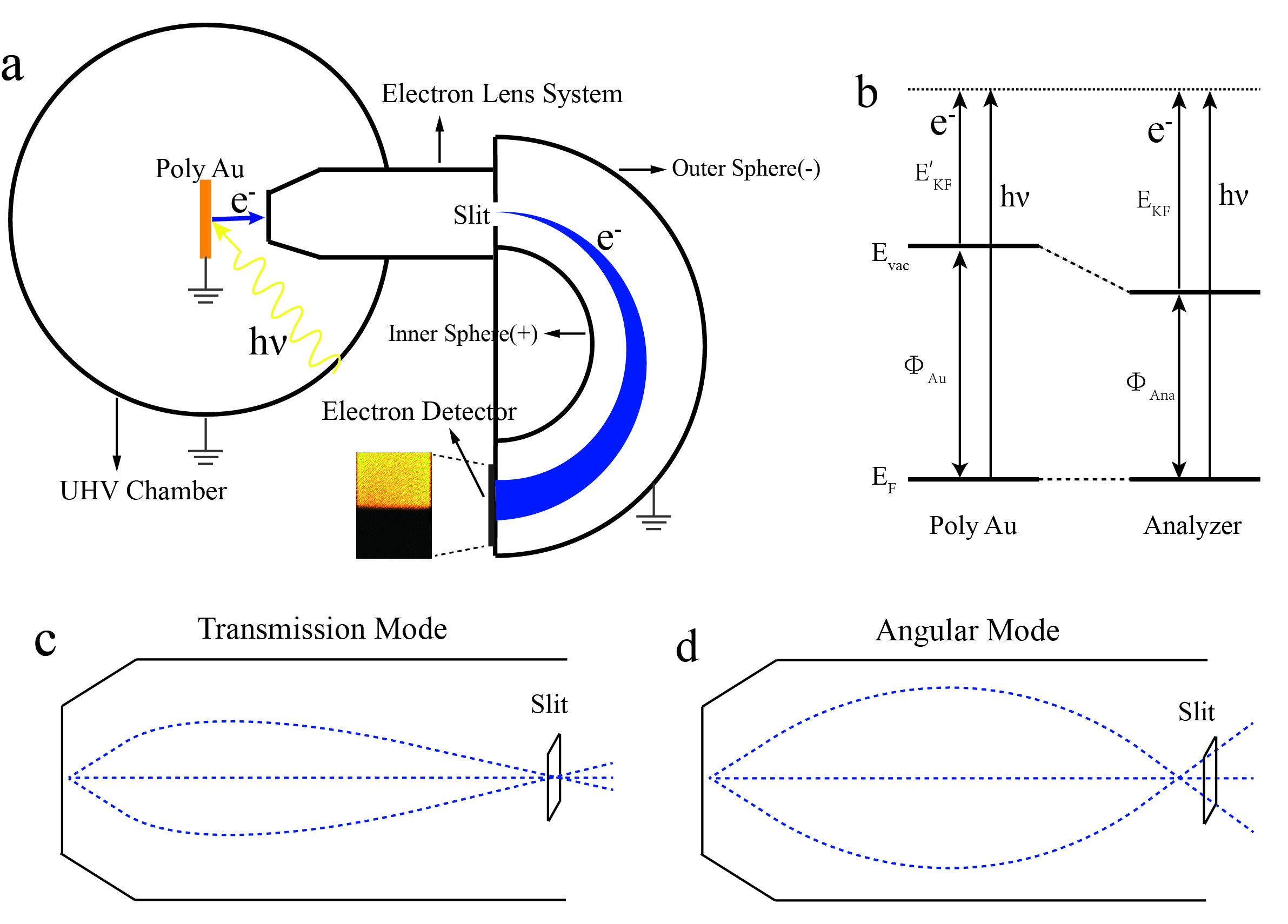}
\end{center}
\caption{{\bf Schematic of the Planck constant measurement with  modern photoemission spectroscopy.} (a) Experimental setup of photoemission spectroscopy with a hemispherical electron energy analyzer. Light with three different wavelengths (and thus photon energies $h\nu$)  can be generated from the helium discharge lamp equipped with a monochromator:  He I$\alpha$ line (58.43339 nm),  He I$\beta$ line (53.70293 nm) and He II$\alpha$ line (30.37858 nm). When the light is incident on a polycrystalline gold (Poly Au),  electrons are emitted out from the sample. The energies and emission angles of the photoemitted electrons are measured by the hemisphere electron energy analyser  with two different modes: angular mode or transmission mode.  (b) The energy levels for the polycrystalline gold and the electron energy analyzer. When the polycrystalline gold and the analyzer are in good electrical contact and grounded, their Fermi levels are lined up at the same energy level. The vacuum level $E_{vac}$ varies with their respective work functions $\Phi_{Au}$ and $\Phi_{Ana}$.  (c) Schematic electron trajectory of the transmission mode inside the electron lens from the top view. In this case, the electrons with different angles emitted from the same spot of the polycrystalline gold are focussed on the same location of the slit and detected on the same position on the detector along the angle direction. (d) Schematic electron trajectory of the angular mode inside the electron lens of the electron energy analyzer from the top view. In this case, electrons with different emission angles will be spread on different positions of the slit and detected on different locations of the detector along the angle direction.
}
\end{figure*}

\begin{figure*}[tbp]
\begin{center}
\includegraphics[width=1.0\columnwidth,angle=0]{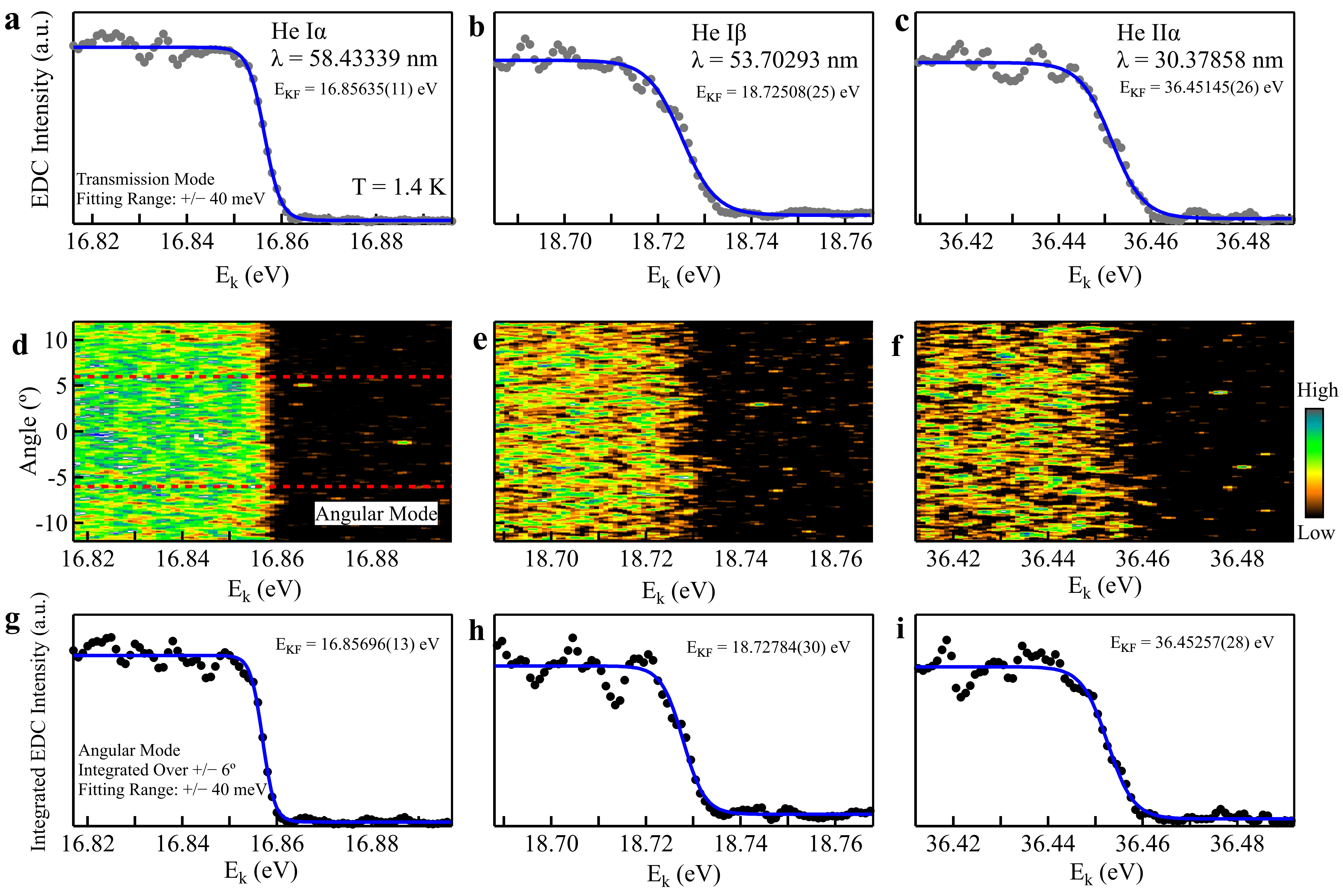}
\end{center}
\caption{{\bf  Measurements of the gold Fermi level using lights with different photon wavelength and different detection modes of the electron energy analyzer.} (a-c) Gold Fermi edge measured at 1.4 K using the transmission mode of the electron energy analyzer under three different light wavelengths:  (a) He I$\alpha$  line (58.43339 nm), (b) He I$\beta$ line (53.70293 nm) and (c) He II$\alpha$ line (30.37858 nm). The energy distribution curve (EDC) is presented which represents the measured photoelectron intensity as a function of the electron kinetic energy ($E_{k}$).  (d-f) Gold Fermi edge measured at 1.4 K using the angular mode of the electron energy analyzer under three different light wavelengths:  (d) He I$\alpha$  line, (e) He I$\beta$ line  and (f) He II$\alpha$ line. Each image represents the measured photoelectron intensity (marked by the false color) as a function of the electron kinetic energy ($E_{k}$, horizontal axis) and the photoelectron emission angle (vertical axis). (g-i) Gold Fermi edge measured at 1.4 K using the angular mode of the electron energy analyzer under three different light wavelengths:  (g) He I$\alpha$  line, (h) He I$\beta$ line  and (i) He II$\alpha$ line. The EDCs are obtained by integrating the photoelectron intensity within an emission angle range of [-6,6] degrees (marked in (d) as two red dashed lines) from the images in (d-f), The  blue solid lines in (a-c) and (g-i) represent fitted curves with Fermi distribution function which determine the energy position of the gold Fermi edge $E_{KF}$.
}
\end{figure*}

\begin{figure*}[tbp]
\begin{center}
\includegraphics[width=1.0\columnwidth,angle=0]{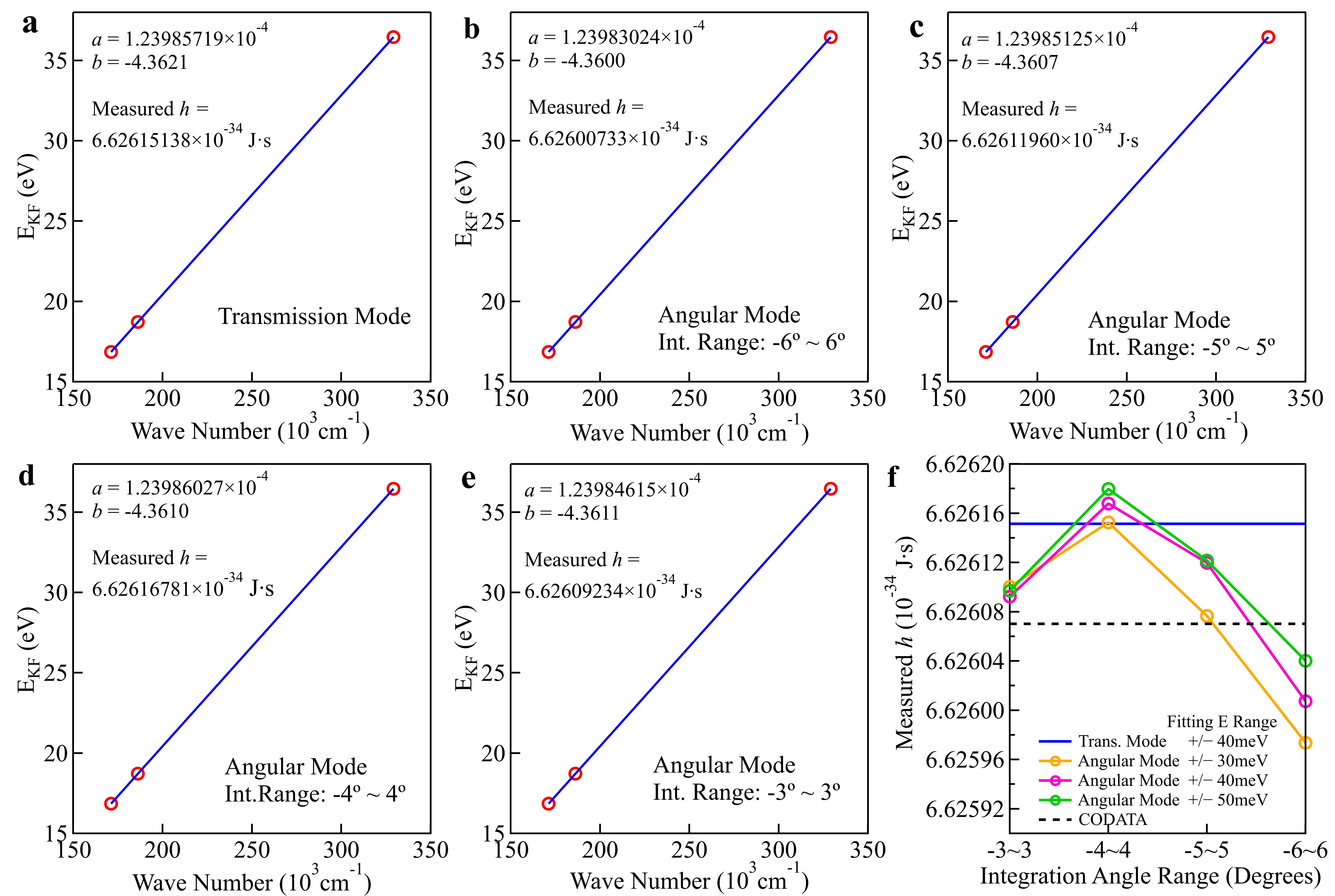}
\end{center}
\caption{{\bf Determination of the Planck constant by linear fitting of the measured Fermi edge energy position at different photon wavelength. } Here $E_{KF}$ in the vertical axis represents the measured energy position of the gold Fermi edge. The wavelength of the light in the horizontal axis is shown in terms of the wave numbers.  The measured energy position of the gold Fermi edge at three different photon wavelengths is linearly-fitted by ($\emph{y} = a\emph{x} + b$). The Planck constant can be obtained from the fitted slope {\it a} while the fitted intercept {\it b} gives the work function of the electron energy analyzer. (a). $E_{KF}$ measured from the transmission mode  as a function of the light wave numbers .  (b-e). $E_{KF}$ measured from the angular mode, integrated within an angle window of [-6,6] degrees (b), [-5,5] degrees (c),  [-4,4] degrees (d) and [-3,3] degrees  (e),  as a function of the light wave numbers . (f) The Planck constant obtained from (a-e).  Different fitting energy ranges of the Fermi-Dirac distribution are also chosen,  [-30,30]meV (yellow), [-40,40]meV (pink) and [-50,50]meV (green) with respect to the Fermi level,  to check its effect on the measured Planck constant values. The black dashed line represents the standard value of the Planck constant by CODATA.
}
\end{figure*}

\begin{figure*}[tbp]
\begin{center}
\includegraphics[width=1.0\columnwidth,angle=0]{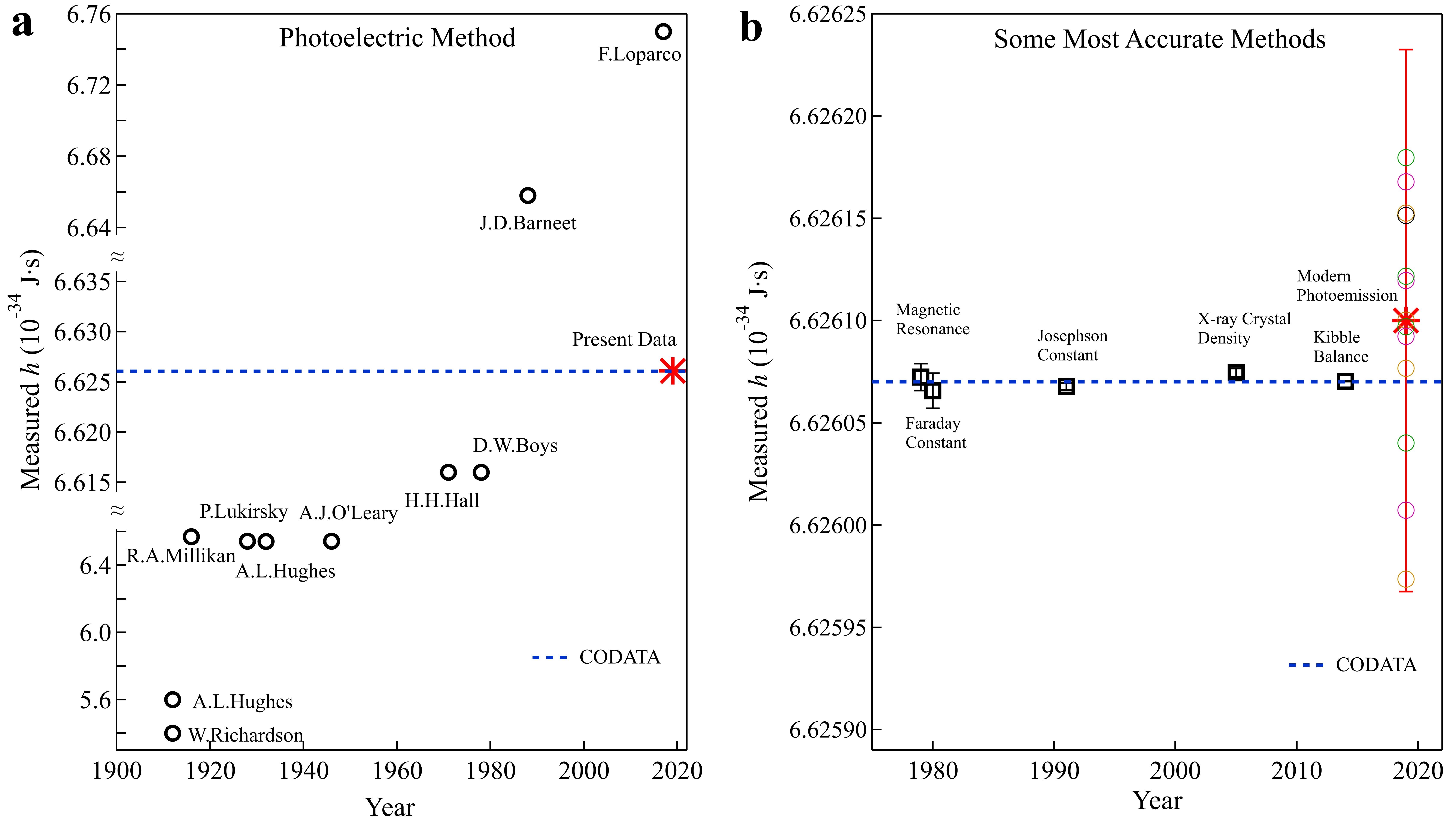}
\end{center}
\caption{{\bf The value of the Planck constant determined by various experimental methods.} (a) The values of the Planck constant measured by the photoelectric effect from different groups at different time.  The red star is our result from modern photoemission spectroscopy. The blue dashed line represents the standard value by CODATA. (b) The values of the Planck constant by the most accurate methods. The empty circles represents the values we obtained from our transmission mode and angular mode measurements in Fig. 3.  The red star is the average value of the Planck constant from the results. The blue dashed line is the standard value of the Planck constant by CODATA.
}

\end{figure*}

\end{document}